# Simultaneous mapping of EMCD signals and crystal orientations in a transmission electron microscope

Hasan Ali[1,3], Jan Rusz[2], Tobias Warnatz[2], Björgvin Hjörvarsson[2], Klaus Leifer[1,*]

[1]Applied Materials Science Division, Department of Engineering Sciences, Uppsala University, Box 534, 75121 Uppsala, Sweden
[2]Department of Physics and Astronomy, Uppsala University, Box 516, 75120 Uppsala, Sweden
[3]Department of Electrical Engineering, Mirpur University of Science and Technology (MUST), Mirpur-10250 (AJK), Pakistan.
[*]Klaus.Leifer@angstrom.uu.se

## Abstract
When magnetic properties are analysed in a TEM using the technique of electron magnetic circular dichroism (EMCD), one of the critical parameters is the sample orientation, and this, independently on the chosen acquisition geometry. Since small orientation changes can have a strong impact on the EMCD measurement, it is experimentally non trivial to measure the EMCD signal as a function of sample orientation. The classical EMCD experimental setup requires to tilt the crystal in a 2-beam orientation and to acquire two electron energy loss spectra at two conjugate scattering angles. The effect of a mistilt from the perfect 2-beam orientation on the measured EMCD signals has not been explored yet due to different experimental constraints. In order to maintain the exact sample location and orientation for the acquisition of the EMCD signal, we have developed a methodology to simultaneously map the quantitative EMCD signals and the local orientation of the crystal. We analyse, both experimentally and with simulations, how the measured magnetic signals evolve with a change in the crystal tilt from the exact 2-beam orientation. Based on this analysis, we establish an accurate relationship between the crystal orientations and the EMCD signals. Our results demonstrate that a small crystal tilt away from the 2-beam orientation can significantly alter the strength and the distribution of the EMCD signals. From an optimisation of the crystal orientation, we obtain quantitative EMCD measurements.

## Introduction

Electron magnetic circular dichroism (EMCD) [1], a transmission electron microscope (TEM) based technique, has emerged as an important technique to determine the magnetic moments of the materials with much higher spatial resolution as compared to its x-ray counterpart XMCD [2]. The EMCD technique was proposed in 2003 [3] and experimentally demonstrated in 2006 [4]. From the time of its discovery, EMCD has seen a continuous rise and researchers have put serious efforts to improve the signal to noise (S/N) ratio [5-9] and the spatial resolution [10-15] of the technique. This led the EMCD technique to explore many materials based problems [16-21] as well as to obtain the magnetic signals from single atomic planes [22, 23].

One of the apparently simple findings in the experimental evolution of the EMCD technique since its discovery is that the sample orientation and the electron beam position must be very well defined in order to obtain a quantitative EMCD signal. The accurate knowledge of sample orientation is thus of uttermost importance when aiming for quantitative atomic resolution EMCD. With this in mind, it is suprising that hitherto, the orientation dependence of the EMCD signal has not been analysed systematically in the experimental situation. Furthermore, when aiming for atomically resolved EMCD work, not only the orientation must be precisely known, but also the position of the electron beam with respect to the exposed atom must be well known and strictly identical for both EELS spectra that are needed for obtaining the EMCD signal. Having acquired the STEM image of the atomic lattice of the magnetic material, the place of the EMCD analysis can be accurately determined. But, it is non-trivial to obtain all information needed for highly accurate EMCD, i.e. orientation and both EELS spectra simultaneously.

In the classical EMCD experimental setup, the TEM sample is tilted to a 2-beam condition (2BC) and two electron energy loss (EELS) spectra are acquired at two conjugate scattering angles. the From an experimental point of view, tilting the TEM sample in a perfect 2BC is not always trivial especially for thin samples and nanoparticles where the Kikuchi lines are not visible. Depending on the thickness and the extinction length of the material, the intensity of the diffracted beam can be lower or even higher than the direct beam. Another difficulty comes when producing the spatial maps of EMCD in scanning TEM (STEM) mode. Considering the fact that most of the crystals are not perfect single crystals and even if they are, the crystal orientation of the TEM sample might locally change within the measured area, thus producing different orientation conditions at different scan points. It has been shown in simulations that a deviation from a 2BC can produce a change in the strength and distribution of the EMCD signals in the reciprocal space but no systematic study has been carried out in this context [24]. One of the major difficulties in doing so is the serial acquisition of the two EELS spectra needed for the EMCD and the diffraction patterns at each beam position which make it hard to ensure the spatial registration among these measurements.

Here we introduce a technique to simultaneously map the EMCD signals and the local orientations of the sample in a single scan of the electron beam. By inserting a custom-made quadruple aperture (QA) [25] in the reciprocal space plane of the electron beam trajectory, we simultaneously obtain four angle-resolved spectroscopic signals in a single acquisition. The four signals include the two magnetic components required for the EMCD measurements and the inelastic intensities of the **0** and the **g** beams. The approach is based on the hypothesis that a ratio of the inelastic intensities of the diffracted ($I'_g$) and the direct ($I'_0$) beams is a quantitative representative of the local orientation of the crystal. With the help of simulations we show how the values of $I'_g/I'_0$ are related to a tilt of the crystal from exact 2-beam orientation. We establish an experimental relationship between the crystal orientation and the EMCD signals and demonstrate that a change in crystal orientation can significantly affect the measured magnetic signals. We find the exact tilt direction in the experiments by analyzing the elastic diffraction patterns acquired immediately after the QA-mapping. We will use the notations $I_0$ and $I_g$ for elastic intensities and $I'_0$ and $I'_g$ for inelastic intensities of the **0** and the **g** beams respectively.

## Experimental Details
We used a single crystal bcc Fe film to demonstrate the experiments. The Fe film was epitaxially grown via direct-current magnetron sputtering in an ultra high vacuum system with a base pressure in the low $10^{-9}$ mbar and an operationg pressure of 2.7 mbar Ar (99.999 99 % purity). Prior to the deposition process, the MgO(001) substrate was annealed for 1 h at 550ºC, while the deposition temperature was kept at 350ºC. Finally, an $Al_2O_3$ film (3 nm thick) was deposited using radio-frequency sputtering at room temperature, helping to prevent oxidation of the underlying Fe layer. X-ray diffraction measurements (not shown) were performed with a Philips X-Pert Pro MRD diffractometer (Cu Kα = 1.5418 Å). A Gaussian fit of the Fe rocking curve peak revealed a full width at half maximum of 1º.

The TEM sample was prepared in a plan-view geometry by mechanical polishing, dimple grinding and Ar- ion milling [26] from the substrate side until perforation appears. In this geometry, the thickness of the Fe film gradually increases from the edge of the hole and reaches 20 nm where it stays the same everywhere. In the following EMCD experiments, the measurements were carried out on an area of the sample where the thickness of the magnetic film (Fe) is 20 nm.

The experiments were performed on a FEI Tecnai-F30 TEM at an acceleration voltage of 300 kV. The instrument is equipped with a Gatan tridiem spectrometer. To set up the experimental conditions, the TEM sample was tilted away from the [001] zone axis to reach a 2-beam condition by exciting **g** = 002 for Fe as shown in FIG. 1 (a). A quadruple aperture (QA) was designed with two bigger holes for EMCD measurements and two smaller holes to map the crystal orientation. In the coming text, we will call the bigger aperture holes "EMCD apertures" and the smaller aperture holes "beam apertures" to avoid confusion. The QA aperture was mounted on the spectrometer entrance aperture (SEA) in an orientation that the four holes of the aperture do not overlap along $q_y$-axis as shown in FIG. 1 (b). The TEM sample was rotated so that the positions of the **0** and **g** beams coincide with the positions of the respective beam apertures. Note that despite the rotation of the QA aperture, the EMCD apertures are located at Thales circle positions with respect to the beam apertures. The semi-collection angle for each EMCD aperture is 3 mrad. For the data acquisition, the TEM was operated in microprobe STEM mode and an electron probe with a semi-convergence angle of 1.6 mrad was scanned across a 100 x 100 nm$^2$ region of the sample as shown by the green square in FIG. 1 (c). The step size between two adjacent scan points was set to 4 nm. A 2D EELS image was acquired at each beam position with a dwell time of 5 s producing a 4D datacube. Due to the distinct positions of the four holes of the QA, the momentum transfer along $q_y$ is preserved for each aperture resulting in four angle resolved spectroscopic traces being projected simultaneously onto the CCD camera (FIG. 1 (d)). This image contains two conjugate EELS spectral traces (here called C+ and C-) produced by the EMCD apertures. At the same time, the inelastic intensities of the **0** ($I'_0$) and **g** ($I'_g$) beams are estimated by summing up the edge intensites of the two spectral traces produced by the beam apertures. The image shown in FIG. 1 (d) is extracted from one scan point of the acquired 4D data cube where the edge intensities produced by the **0** and the **g** beam are very close to each other. The C+ and C- EELS spectra are extracted from the regions marked by blue and red rectangles on the uppermost and the lowermost spectral traces respectively. The background of the EELS spectra is subtracted by fitting a power law model and extrapolating it under the energy loss edges. After that the post edge of the two spectra is normalized and the EMCD signal is obtained by taking the difference of C+ and C- spectra as shown in FIG. 1 (e). The resulting difference signal shows the EMCD signature with the inverse signs at the L$_3$ and somewhat weaker at the L$_2$ energy loss edges of Fe.

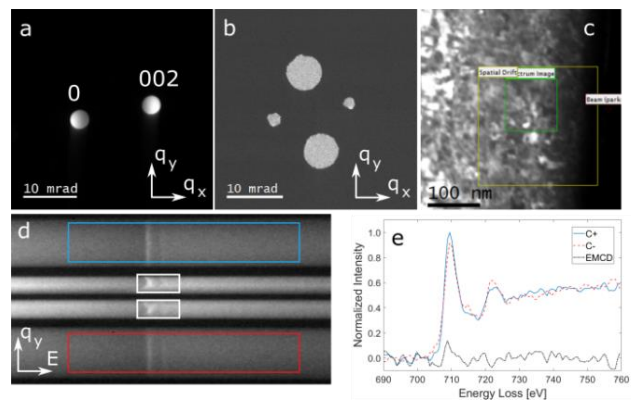

**FIG. 1. (a) Experimentally obtained 2-beam diffraction condition for EMCD experiments with g=002 as diffracted beam. (b) An image of the quadruple aperture. (c) ADF survey image used to acquire the EMCD-orientation maps**

with green box showing the measurements area. (d) 2D EELS image extracted from one scan point of the acquired 4D data cube. The C+ and C- EELS spectra for the EMCD measurements were extracted from the upper and lower traces respectively by integrating the intensities in the marked regions whereas the intensities of g and 0 beams were determined by summing up the intensities in the white rectangles marked in the second and third traces respectively. (e) The background subtracted and post edge normalized EELS spectra extracted from (d) together with their difference (EMCD) signal.

To produce the maps shown in the results section, MATLAB scripts are written and utilized together with digital micrograph (DMG). In the first step, the energy drift is corrected by aligning the peaks in the 2D EELS images located at different scan points in the 4D dataset. After that, a MATLAB script is used to generate two 3D EELS spectrum images (SI) by extracting the EELS spectra from the uppermost and the lowermost spectral traces in the 2D EELS images at each scan point. The background of each EELS SI is subtracted by fitting and extrapolating a power law background model with a fit window 640-700 eV. The post edge of the EELS spectra situated at each pixel of the background-subtracted SIs is normalized by using a normalization window 750-790 eV. The resulting datasets are here called C+ and C- EELS SIs. The $L_3$ and $L_2$ difference maps are produced by fitting an energy window centred at the maxima and calculating the intensity difference between $L_3$ and $L_2$ edges of the EELS spectra at each corresponding pixel of the C+ and C- SIs. Another MATLAB script calculates $I'_g/I'_0$ ratio at each scan point of the 4D dataset by extracting the intensities of the **0** and the **g** beams in the white rectangles marked in FIG. 1 (d). The $I'_g/I'_0$ map is then generated by putting the values back at the corresponding pixels in a 2D matrix.

## Simulations

To verify the hypothesis that the values of $I'_g/I'_0$ at each beam position are representative of the specimen's orientation we have performed simulations of diffraction patterns for both elastic and inelastic electron scattering of a parallel beam on an iron sample tilted into a systematic row orientation, while varying the Laue circle center.

Computational details were set in the following way. Beam acceleration voltage was set to 300kV, as in the experiment. The sample was tilted by approximately 10 degrees from the (001) zone axis into the systematic row orientation. This brings the sample into the three-beam orientation with the same strength of excitation of both **+g** and **–g** = (200) beams. On top of that we introduced a tilt parallel to the systematic row, varying within a range from -6 mrad to +20 mrad with a step of 1 mrad, i.e., in total we have calculated diffraction patterns for 27 different orientations. Sample thicknesses up to approximately 30 nm were considered, but in the figures below we focus on the thickness range from 15 nm to 25 nm, centred on the experimental sample thickness of 20 nm. Calculations were performed using MATS.v2 [27] code, modified to also output the relevant information from the elastic scattering of the incoming beam. We have calculated diffraction patterns for inelastic scattering of electrons within the Fe $L_3$-edge energy range and $\theta_x$ = (-2 mrad, 16 mrad) and $\theta_y$ = (-12 mrad, 12 mrad) ranges of the scattering angles. MATS.v2 convergence parameter was set to 2 x $10^{-5}$, well sufficient for a plane-wave calculation.

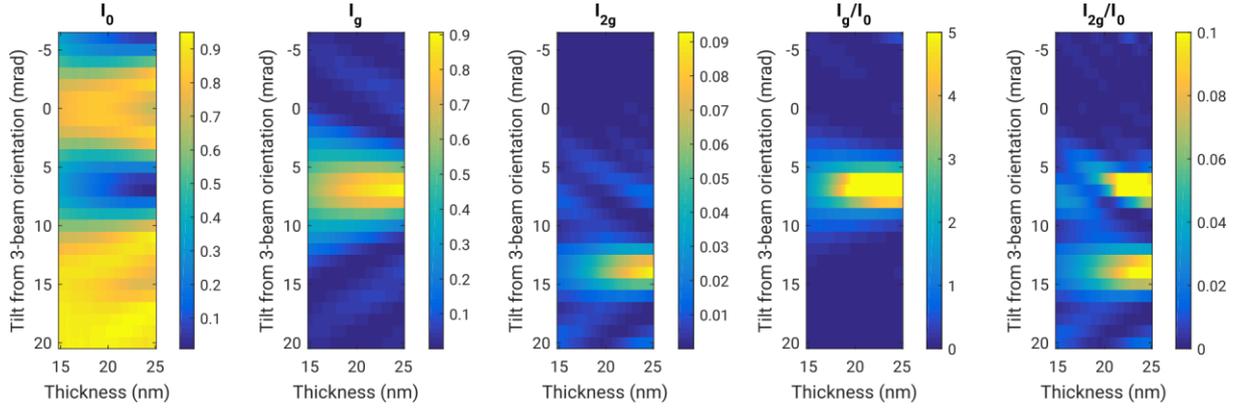

FIG. 2. Results from elastic scattering calculations of the intensities of the transmitted beam and the Bragg scattered beams g = (200) and 2g as a function of tilt from three-beam orientation and sample thickness for bcc iron crystal at 300kV acceleration voltage.

First we summarize results of the calculations of the elastic scattering. In the experiments, a diffraction mapping using elastically scattered electrons was performed over approximately the same area on which the map containing the spectroscopic information was taken. From this dataset, the intensities of $I_0$, $I_g$ and $I_{2g}$ were extracted. FIG. 2 shows the elastic scattering calculation results of intensities of these beams and their ratios as a function of sample thickness and the tilt from the 3-beam orientation. The most prominent feature is the strong reduction of the intensity of the transmitted beam at around 7 mrad beam tilt, especially at a thickness approaching 25 nm. This tilt is close to the exact 2-beam orientation i.e. where the excitation error of the **g** beam is close to zero. (Note a similar feature developing towards a beam tilt of -7 mrad, when it would be the -**g** beam approaching low excitation error.) Simultaneously, the intensity of the **g** beam increases. This is an example of well-known Pendellösung oscillations. At approximately 14 mrad tilt, the 2**g** beam approaches zero excitation error and its intensity grows in that region. However, this beam has a significantly higher extinction distance and therefore at thicknesses below 25 nm it doesn't develop significant intensity. The ratio of the **g**-beam and the transmitted beam intensities highlights the intensity oscillation even more clearly, reaching very high values near the exact two-beam orientation, especially for higher sample thicknesses where the sample thickness gets closer to the extinction length. The ratio $I_{2g}/I_0$ shows two local maxima, one near the exact two-beam orientation, which is due to the afore-mentioned minimum of the transmitted beam intensity, and the second near the tilt that minimizes the excitation error of the 2**g** beam – bringing thus its intensity up.

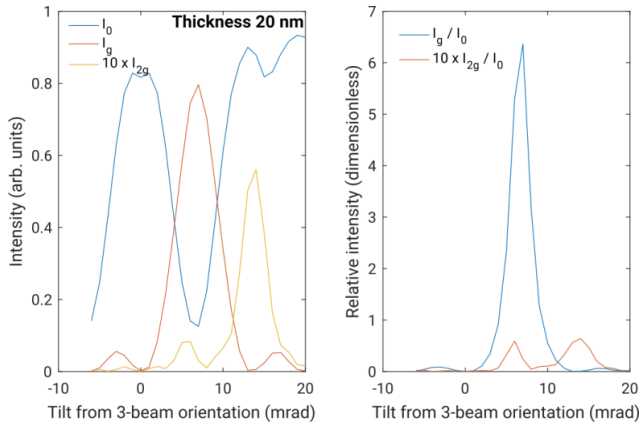

FIG. 3. Intensities of transmitted beam and Bragg-scattered beams g and 2g and their ratios as a function of beam tilt for a 20 nm thick sample of bcc iron.

A more detailed insight, particularly relevant for the interpretation of experiments conducted in this work, are the results at sample thickness of 20 nm. FIG. 3 shows line profiles with beam intensities and their ratios as a function of the tilt. The intensity of **g** beam $I_g$ is well above the intensity of the transmitted beam $I_0$, when the sample is close to an exact two-beam orientation. The intensity ratio $I_g/I_0$ peaks at a value of more than 6. On the other hand the 2**g** beam (due to its large extinction distance) has very low intensities across the whole range of the tilts (note that its intensity was scaled up by factor of 10 for better visibility in the FIG. 3). At any orientation, its relative intensity does not go above approximately 5-6 % of the transmitted beam intensity at this sample thickness.

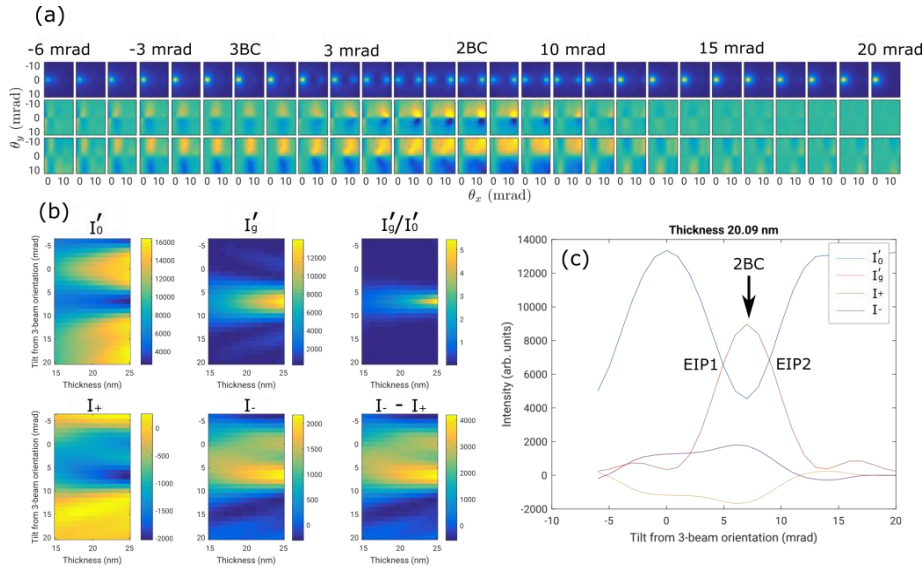

FIG. 4. (a) non-magnetic (top row), magnetic (EMCD; middle row) and relative magnetic (bottom row) diffraction patterns for calculated tilts from the three-beam orientation at sample thickness of 20 nm (b) inelastic intensities of 0 and g beam and their ratio as a function of sample thickness and tilt from 3BC (c) inelastic intensities of 0 and g beam and the magnetic signals plotted as a function of tilt from 3-beam orientation.

We proceed with the results of inelastic scattering calculations. FIG. 4 (a) shows the non-magnetic (top row), magnetic (EMCD; middle row) and relative magnetic (bottom row) diffraction patterns for all calculated tilts from the three-beam orientation at sample thickness of 20 nm. Note, how the area with the strongest magnetic signal moves throughout the diffraction plane. In absolute strength it is largest in the vicinity of the **g** beam, although there it is superposed on tails of the large non-magnetic signal. For that reason the relative strength of the magnetic signal is not necessarily optimal there. Overall, the relative strength of the magnetic signal seems to be rather stable near the Thales circle detector positions, when the beam tilt is near the two-beam orientation.

These diffraction patterns were further analysed by placing virtual detector apertures as in the experiment. That effectively reduces the diffraction patterns down to four number per beam tilt, representing the non-magnetic intensities of inelastically scattered electrons at the transmitted beam and **g** beam and the magnetic intensities at (+) and (-) detector positions. FIG. 4 (b) summarizes these results as a function of tilt and sample thickness, including the ratio of intensities at the transmitted beam and the **g** beam. Qualitatively the results for non-magnetic components of the inelastic scattering cross-section strongly remind the results of the elastic calculations shown in FIG. 2. This is due to preservation of the elastic contrast, as discussed elsewhere [28-31].

Reducing the data further, we focus in FIG. 4 (c) on the four inelastic intensities at a sample thickness of 20 nm. Again, the non-magnetic results are qualitatively similar to the results from elastic calculations shown in FIG. 3, however there are quantitative differences. For example, the **g** beam intensity only reaches about 60% of the maximal intensity of the transmitted beam, while in the elastic calculation it was up to about 90%. These quantitative differences might be related to the limited small detector aperture used to collect the inelastic scattering intensities (only 1 mrad semi-angle). However, we have not explored this effect further.

In view of the experimental results on EMCD, shown below, we note that in both the elastic and inelastic calculations, equal intensities of the transmitted beam and the **g**-beam are not obtained at the exact 2-beam orientation for the sample thickness of 20 nm. Instead, the equal intensities are obtained at two different tilts, at approximately 2 – 2.5 mrad away from the exact two beam orientation. For simplicity, we will refer these two points as 'equal intensity points (EIP)' in the coming text. In addition, we point out that the EMCD strength (both absolute and relative) is not maximal, when the intensities of the transmitted and **g** beam are equal.

## Experimental Results

To verify the simulations shown in FIG. 4, we calculated the ratio of inelastic intensities of the **0** and the **g** beams ($I'_g/I'_0$) and the difference signals at the $L_3$ and $L_2$ energy loss edges from the spectroscopic images at each scan point of the 4D dataset. The results are presented in FIG. 5 where the three maps represent the same area of the sample which was scanned by the electron beam. The $I'_g/I'_0$ map shows a region of similar orientation within the measured area which is oriented close to 2-beam condition (2BC) whereas the orientation of the regions surrounding the well-oriented grain is quite far away from the 2BC. It is interesting to note that we obtain the maximum value of $I'_g/I'_0 = 1$ so even the best orientation within the scanned region is 2-2.5 mrad away from the exact 2BC and it can be at one of the two equal intensity points (EIPs) as shown in FIG. 4 (c). Continuing, a clear correlation is observed among the three maps. The well-oriented region of the sample results in large positive values of $L_3$ difference signal and negative values of $L_2$ difference signal as expected by the EMCD signature. The $L_3$ and $L_2$ difference signals, on the other hand, weaken or even change their sign in the misoriented regions. Thus it seems that a change in crystal orientation away from the 2BC may significantly influence the measured EMCD signals and we have also seen it in FIG. 4 (a) where a crystal tilt from 2BC results in substantial variations in the EMCD signal strength and distribution within the diffraction plane.

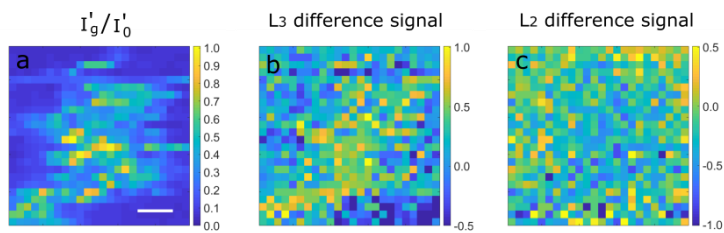

**FIG. 5. The maps of (a) $I_g/I_0$ (b) $L_3$ difference and (c) $L_2$ difference values. A clear correlation can be seen among the maps where the areas with higher $I_g/I_0$ values result in positive difference at $L_3$ and negative difference at $L_2$. The scale bar is 20 nm.**

To more closely observe the variations in $L_3$ and $L_2$ difference signals with a corresponding change in the value of $I'_g/I'_0$, we plotted the $L_{3,2}$ difference values against the $I'_g/I'_0$ for all the pixels with $I'_g/I'_0$ ordered from the smallest to the largest values. The plots are presented in FIG. 6 where the experimental data is plotted as scatter plots. A running average of the experimental data with an interval of 0.05 is obtained and is plotted as solid lines overlaid on the experimental data. Again looking at the FIG. 4 (c), the value of $I'_g/I'_0$ reduces to 0.1 with approximately 5 mrad tilt from each of the EIP with the tilt direction away from the 2BC. So, the tilt range of $I'_g/I'_0$ axis in FIG. 6 can be considered 5 mrad from 0.1-1.0 with 1.0 representing a zero tilt from one of the EIP. All the values less than 0.1 can represent a tilt more than 5 mrad up to as high as approximately 15 mrad.

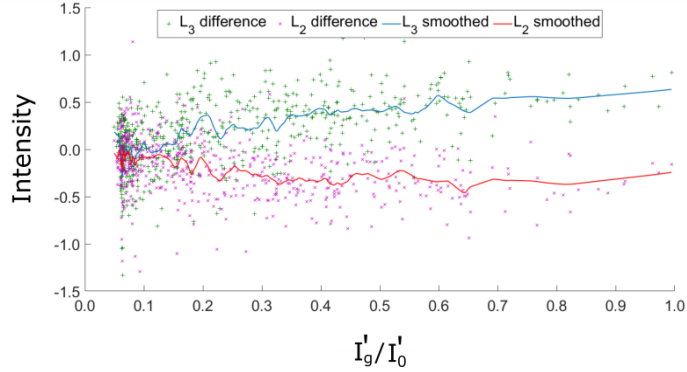

FIG. 6. The difference values at $L_3$ and $L_2$ edges plotted against the $I_g/I_0$. The scatter plots are the experimental values whereas the solid curves are the running average of the experimental data over an interval of 0.05.

Qualitatively comparing the simulated magnetic signal curves in FIG. 4 (c) with the experimental curves in FIG. 6, there is an impression that the curves match more closely considering a tilt towards 3 beam orientation from EIP1. To investigate the exact tilt direction and EIP in our experiment, we acquired a 4D diffraction scan dataset from nearly the same region measured in the above experiment. This dataset contains an elastic diffraction pattern at each scan point. The dataset was acquired immediately after the QA-scan without moving the diffraction pattern and in the already set conditions we missed the –**g** beam on the CCD camera but 2**g** (004 for Fe) beam was captured in the diffraction patterns. The intensity changes in 2**g** can be used to find out the tilt direction. We calculated the intensities of $I_g/I_0$ and $I_{2g}/I_0$ from the diffraction scan dataset and plotted them together in an order of the smallest to the largest values of $I_g/I_0$ as shown in FIG. 7. According to the simulations shown in FIG. 3, considering EIP2 as the point of equal intensities, a tilt away from the 2BC causes an increase in $I_{2g}/I_0$ with a decrease in $I_g/I_0$. On the other hand, if we consider the EIP1, $I_{2g}/I_0$ does not show any significant change and stays to very low values with a decrease in $I_g/I_0$. The experimental intensity plots presented in FIG. 7 clearly match the latter case confirming that the equal intensities of $I_0$ and $I_g$ obtained in our experiments represent EIP1 and the tilt direction within the scanned area is one corresponding to an orientation with the sample tilted from the 2BC towards the 3BC orientation. This also confirms our qualitative observation made above that the experimental magnetic signal curves match to the simulated magnetic curves in a tilt direction towards the 3-beam orientation.

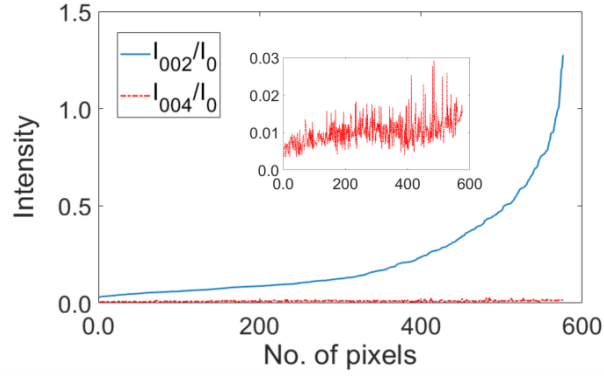

**FIG. 7.** The quantities $I_{002}/I_0$ and $I_{004}/I_0$ are plotted together in an order from the lowest to the highest values of $I_{002}/I_0$. The intensities are extracted from a 4D diffraction data cube.

To investigate the effects of a change in the crystal orientation on the resulting EMCD spectra and the measured magnetic properties, we produced the EMCD signals for different values of $I'_g/I'_0$. For this purpose, we divided the data into four $I'_g/I'_0$ intervals of 0.76-1.0, 0.50-0.75, 0.26-0.50 and 0.0-0.25 and summed up all the EELS spectra within each interval to produce single C+ and C- EELS spectra for each interval. The background of the EELS spectra was subtracted using a power law model and the spectra were post edge normalized using the same window sizes as described in the methods section. At this stage of post-processing, an additional difficulty is encountered which arises due to the position of the EMCD apertures at different xy-coordinates on the CCD camera. Focussing the EELS edges for both apertures during the experiment is not trivial and there are still some distortions left in the lower spectral trace. These distortions result in blurring the energy loss edges causing the edges in the lower spectral trace to be a little broader than the upper spectral trace. Here we adopted the profile matching routine recently introduced and used in Ref. [32] to sharpen the edges in the lower trace keeping the area under the edge same. After profile matching, the difference of the two spectra is taken which is the EMCD signal. The EMCD signals obtained for different $I'_g/I'_0$ intervals are shown in FIG. 8. A clear EMCD signal at both the $L_3$ and $L_2$ edges can be observed in all the cases. The strength of the EMCD signal at $L_3$ edge goes down from 8.3 % to 4 % from the best to the worst orientation case whereas for the $L_2$ edge, it goes down from 6.3 % to 3.8 %.

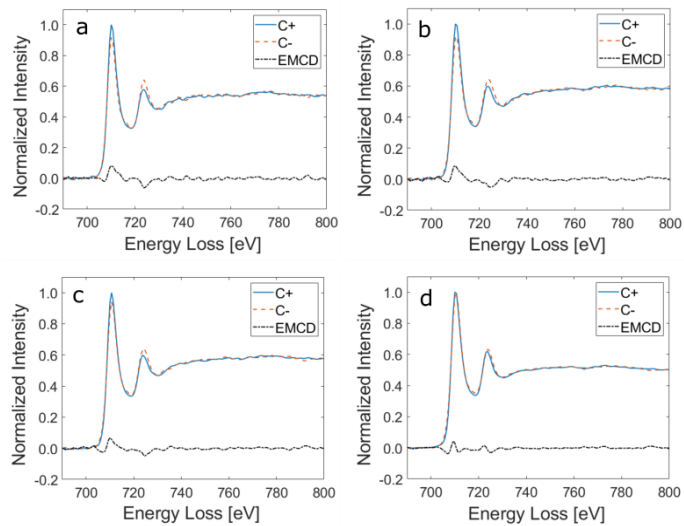

**FIG. 8.** EMCD spectra extracted from the map with ($I_g/I_0$) orientation intervals (a) 0.76-1.0 (b) 0.51-0.75 (c) 0.26-0.50 (d) 0.0-0.25. The total number of spectra summed up for different intervals are different.

We calculated the $m_L/m_S$ values of the four EMCD spectra shown above. The intensity difference at $L_3$ and $L_2$ energy loss edges was calculated by fitting a 4 eV window centred at the maxima of the edges and integrating the intensities within the window. The values were then put in the following formula to calculate the $m_L/m_S$ values.

$$\frac{m_L}{m_S} = \frac{2}{3} \frac{\int_{L3} \Delta I(E)dE + \int_{L2} \Delta I(E)dE}{\int_{L3} \Delta I(E)dE - 2\int_{L2} \Delta I(E)dE} \quad (1)$$

The results are summarized in Table 1. The $m_L/m_S$ values obtained for the first three cases are similar within the error bars and are also in good agreement with the previously reported values for bcc Fe [5, 33-35].

Table 1. The calculated values of $m_L/m_S$ from the EMCD spectra obtained for different intervals of $I_g/I_0$.

| $I_g/I_0$ range | Intensity difference at $L_3$ | Intensity difference at $L_2$ | Calculated $m_L/m_S$ |
|---|---|---|---|
| 0.76-1.0 | 0.36 | -0.27 | 0.06 ± 0.01 |
| 0.51-0.75 | 0.37 | -0.28 | 0.06 ± 0.01 |
| 0.26-0.50 | 0.26 | -0.21 | 0.05 ± 0.01 |
| 0.0-0.25 | 0.029 | -0.072 | -0.16 ± 0.01 |

For the last case, the $L_2$ difference signal becomes higher than the $L_3$ difference signal, resulting in a negative value of $m_L/m_S$. According to simulations shown in FIG. 4 (c), the range $I_g/I_0$=0.0-0.25 corresponds to crystal orientations near or beyond the 3BC. The bad values of $m_L/m_S$ for this range might be an effect of asymmetry [36] which becomes dominant for larger tilts. Nevertheless, the EMCD signal strength is relatively stable and the measured $m_L/m_S$ values are not affected within a tilt range of little more than 5 mrad. These results match closely to the simulated results for the tilt direction towards the 3-beam orientation.

## Discussion

According to the simulations (FIG. 4), the EMCD strength for a 20 nm thick iron sample is maximized at tilts 6-7 mrad from the three-beam orientation (7 mrad is almost precisely the two-beam orientation). The signal strength reduces with a tilt away from the 2BC but the reduction is not symmetric in the two tilt directions. The EMCD signal rapidly decreases for a tilt towards EIP2 and becomes 0 within a tilt range of 5 mrad (from 7 mrad to 12 mrad), after which it reverses its sign. On the other hand, the EMCD signal stays relatively stable within a tilt range of 10 mrad for a tilt towards the 3 beam orientation and decreases gradually. Although these results are system and orientation dependent, they can give some qualitative hints, why the EMCD strength was often very weak in actual measurements, in particular $L_2$ EMCD signals that were often at the limit of detection or had even disappeared. Apart from measuring a sample with unsuitable thickness (close to extinction distance of the **g** beam) it could be due to setting an orientation with equal intensities of the **0** and **g** beams, tilting too far away from the two-beam orientation. From our results it appears to be safer to first set a three-beam orientation and then quantitatively tilt further towards an exact two-beam orientation (regardless of the relative intensities of the **0** and **g** beam). When the quantitative tilting is not possible or not practical, it is advisable to tilt slowly until the first orientation with equal intensities of **0** and **g** beams is reached and measure there. In an optimal case one reaches the exact two-beam orientation, or a slightly smaller tilt, however, the EMCD strength seems to be relatively robust for tilts few mrad smaller than the exact two-beam orientation. This is confirmed by the experimentally obtained EMCD signals with different crystal orientations lying in between the two and the three beam orientations (FIG. 8).

## Conclusions

We have developed a methodology to simultaneously acquire the EMCD signals and the local crystal orientations in a single electron beam scan in the TEM. This ensures the determination of the effects of crystal tilt on the measured EMCD signals with high accuracy. The experimental results and the simulations show that the EMCD signal strength weakens as the crystal tilts away from the exact 2-beam orientation. The change is not symmetric for the plus and minus tilts from the 2BC and the EMCD signals stay more stable in a tilt range from the 2-beam towards the 3-beam orientation. We quantitatively show that the measured $m_L/m_S$ values stay stable in this safe tilt range but can significantly alter otherwise. The work shows the significance of setting up the better orientation conditions to obtain good EMCD signals. The experimental setup can be utilized for high precision quantitative EMCD measurements for any magnetic crystal tilted in 2-beam orientation.

## Acknowledgements

We acknowledge Eric Lindholm for fabricating the apertures. K.L. gratefully acknowledge the support from the Swedish Science Council, VR grants number 2016 05259. J.R. acknowledges Swedish Research Council for financial support. The simulations were performed on resources provided by the Swedish National Infrastructure for Computing (SNIC) at the National Supercomputer Centre at Linköping University (NSC).

## References

[1] D. S. Song, Z. Q. Wang, and J. Zhu, "Magnetic measurement by electron magnetic circular dichroism in the transmission electron microscope," *Ultramicroscopy,* vol. 201, pp. 1-17, Jun 2019.

[2] T. Funk, A. Deb, S. J. George, H. Wang, and S. P. Cramer, "X-ray magnetic circular dichroism— a high energy probe of magnetic properties," *Coordination Chemistry Reviews,* vol. 249, pp. 3-30, 2005.

[3] C. Hébert and P. Schattschneider, "A proposal for dichroic experiments in the electron microscope," *Ultramicroscopy,* vol. 96, pp. 463-8, 2003.

[4] P. Schattschneider, S. Rubino, C. Hébert, J. Rusz, J. Kuneš, P. Novák, E. Carlino, M. Fabrizioli, G. Panaccione, and G. Rossi, "Detection of magnetic circular dichroism using a transmission electron microscope," *Nature,* vol. 441, pp. 486-488, 2006.

[5] H. Lidbaum, J. Rusz, A. Liebig, B. Hjörvarsson, P. M. Oppeneer, E. Coronel, O. Eriksson, and K. Leifer, "Quantitative magnetic information from reciprocal space maps in transmission electron microscopy," *Physical Review Letters,* vol. 102, pp. 1-4, 2009.

[6] K. Tatsumi, S. Muto, J. n. Rusz, T. Kudo, and S. Arai, "Signal enhancement of electron magnetic circular dichroism by ultra-high-voltage TEM, toward quantitative nano-magnetism measurements," *Microscopy,* vol. 63, pp. 243-247, 2014.

[7] J. Verbeeck, C. Hebert, S. Rubino, P. Novak, J. Rusz, F. Houdellier, C. Gatel, and P. Schattschneider, "Optimal aperture sizes and positions for EMCD experiments," *Ultramicroscopy,* vol. 108, pp. 865-872, Aug 2008.

[8]  Z. C. Wang, X. Y. Zhong, L. Jin, X. F. Chen, Y. Moritomo, and J. Mayer, "Effects of dynamic diffraction conditions on magnetic parameter determination in a double perovskite Sr2FeMoO6 using electron energy-loss magnetic chiral dichroism," *Ultramicroscopy,* vol. 176, pp. 212-217, 2017.

[9]  H. Ali, T. Warnatz, L. Xie, B. Hjörvarsson, and K. Leifer, "Quantitative EMCD by use of a double aperture for simultaneous acquisition of EELS," *Ultramicroscopy,* vol. 196, pp. 192-196, 2019.

[10]  P. Schattschneider, C. Hébert, S. Rubino, M. Stöger-Pollach, J. Rusz, and P. Novák, "Magnetic circular dichroism in EELS: Towards 10 nm resolution," *Ultramicroscopy,* vol. 108, pp. 433-438, 2008.

[11]  S. Schneider, D. Pohl, S. Löffler, J. Rusz, D. Kasinathan, P. Schattschneider, L. Schultz, and B. Rellinghaus, "Magnetic properties of single nanomagnets: Electron energy-loss magnetic chiral dichroism on FePt nanoparticles," *Ultramicroscopy,* vol. 171, pp. 186-194, 2016.

[12]  P. Schattschneider, M. Stöger-Pollach, S. Rubino, M. Sperl, C. Hurm, J. Zweck, and J. Rusz, "Detection of magnetic circular dichroism on the two-nanometer scale," *Physical Review B,* vol. 78, pp. 104413-104413, 2008.

[13]  Z. Q. Wang, X. Y. Zhong, R. Yu, Z. Y. Cheng, and J. Zhu, "Quantitative experimental determination of site-specific magnetic structures by transmitted electrons," *Nature Communications,* vol. 4, pp. 1395-1395, 2013.

[14]  T. Thersleff, J. Rusz, B. Hjörvarsson, and K. Leifer, "Detection of magnetic circular dichroism with subnanometer convergent electron beams," *Physical Review B,* vol. 94, pp. 134430-134430, 2016.

[15]  T. Thersleff, S. Muto, M. Werwiński, J. Spiegelberg, Y. Kvashnin, B. Hjörvarsson, O. Eriksson, J. Rusz, and K. Leifer, "Towards sub-nanometer real-space observation of spin and orbital magnetism at the Fe/MgO interface," *Scientific Reports,* vol. 7, pp. 44802-44802, 2017.

[16]  T. Thersleff, J. Rusz, S. Rubino, B. Hjörvarsson, Y. Ito, N. J. Zaluzec, and K. Leifer, "Quantitative analysis of magnetic spin and orbital moments from an oxidized iron (1 1 0) surface using electron magnetic circular dichroism," *Scientific Reports,* vol. 5, pp. 13012-13012, 2015.

[17]  X. Fu, B. Warot-Fonrose, R. Arras, D. Demaille, M. Eddrief, V. Etgens, and V. Serin, "Energy-loss magnetic chiral dichroism study of epitaxial MnAs film on GaAs(001)," *Citation: Appl. Phys. Lett,* vol. 107, pp. 62402-62402, 2015.

[18]  X. Fu, B. Warot-Fonrose, R. Arras, G. Seine, D. Demaille, M. Eddrief, V. Etgens, and V. Serin, "In situ observation of ferromagnetic order breaking in MnAs/GaAs(001) and magnetocrystalline anisotropy of α -MnAs by electron magnetic chiral dichroism," *Physical Review B,* vol. 93, pp. 104410-104410, 2016.

[19]  G. Li, D. Song, Z. Peng Li, and J. Zhu, "Determination of magnetic parameters in La 0.7 Sr 0.3 MnO 3 /SrTiO 3 thin films using EMCD," *Appl. Phys. Lett. Applied Physics Letters,* vol. 108, 2016.

[20]  D. Song, L. Ma, S. Zhou, and J. Zhu, "Oxygen deficiency induced deterioration in microstructure and magnetic properties at $Y_3Fe_5O_{12}$/Pt interface," *Applied Physics Letters,* vol. 107, pp. 042401-042401, 2015.

[21]  Z. H. Zhang, H. L. Tao, M. He, and Q. Li, "Origination of electron magnetic chiral dichroism in cobalt-doped ZnO dilute magnetic semiconductors," *Scripta Materialia,* vol. 65, pp. 367-370, 2011.

[22]  J. Rusz, S. Muto, J. Spiegelberg, R. Adam, K. Tatsumi, D. E. Bürgler, P. M. Oppeneer, and C. M. Schneider, "Magnetic measurements with atomic-plane resolution," *Nature Communications,* vol. 7, pp. 12672-12672, 2016.

[23]  Z. Wang, A. H. Tavabi, L. Jin, J. Rusz, D. Tyutyunnikov, H. Jiang, Y. Moritomo, J. Mayer, R. E. Dunin-Borkowski, R. Yu, J. Zhu, and X. Zhong, "Atomic scale imaging of magnetic circular dichroism by achromatic electron microscopy," *Nature Materials*.

[24]  H. Lidbaum, J. N. Rusz, S. Rubino, A. Liebig, B. H. Orvarsson, P. M. Oppeneer, O. Eriksson, and K. Leifer, "Reciprocal and real space maps for EMCD experiments," *Ultramicroscopy,* vol. 110, pp. 1380-1389, 2010.

[25]   H. Ali, T. Warnatz, L. Xie, B. Hjörvarsson, and K. Leifer, "Towards Quantitative Nanomagnetism in Transmission Electron Microscope by the Use of Patterned Apertures," *Microscopy and Microanalysis,* vol. 25, pp. 654-655, 2019.

[26]   R. Alani and P. R. Swann, "PRECISION ION POLISHING SYSTEM - A NEW INSTRUMENT FOR TEM SPECIMEN PREPARATION OF MATERIALS," *Materials Research Society,* vol. 254, pp. 43-63, 1992.

[27]   J. Rusz, "Modified automatic term selection v2: A faster algorithm to calculate inelastic scattering cross-sections," *Ultramicroscopy,* vol. 177, pp. 20-25, Jun 2017.

[28]   A. Bakenfelder, I. Fromm, L. Reimer, and R. Rennekamp, "Contrast in the electron spectroscopic imaging mode of a TEM," *Journal of Microscopy,* vol. 159, pp. 161-177, 1990.

[29]   A. Howie, "Inelastic scattering of electrons by crystals. I. The theory of small-angle in elastic scattering," *Proceedings of the Royal Society of London. Series A. Mathematical and Physical Sciences,* vol. 271, pp. 268-287, 1963.

[30]   H. G. Brown, A. J. D'Alfonso, B. D. Forbes, and L. J. Allen, "Addressing preservation of elastic contrast in energy-filtered transmission electron microscopy," *Ultramicroscopy,* vol. 160, pp. 90-97, 2016.

[31]   S. L. Cundy, A. Howie, and U. Valdrè, "Preservation of electron microscope image contrast after inelastic scattering," *Philosophical Magazine,* vol. 20, pp. 147-163, 1969.

[32]   T. Thersleff, L. Schönström, C.-W. Tai, R. Adam, D. E. Bürgler, C. M. Schneider, S. Muto, and J. Rusz, "Single-pass STEM-EMCD on a zone axis using a patterned aperture: progress in experimental and data treatment methods," *Scientific Reports,* vol. 9, pp. 18170-18170, 2019.

[33]   C. T. Chen, Y. U. Idzerda, H. J. Lin, N. V. Smith, G. Meigs, E. Chaban, G. H. Ho, E. Pellegrin, and F. Sette, "Experimental Confirmation of the X-Ray Magnetic Circular Dichroism Sum Rules for Iron and Cobalt," *Physical Review Letters,* vol. 75, pp. 152-155, 1995.

[34]   L. Calmels, F. Houdellier, B. Warot-Fonrose, C. Gatel, M. J. Hÿtch, V. Serin, E. Snoeck, and P. Schattschneider, "Experimental application of sum rules for electron energy loss magnetic chiral dichroism," *Physical Review B,* vol. 76, pp. 060409-060409, 2007.

[35]   B. Warot-Fonrose, C. Gatel, L. Calmels, V. Serin, and P. Schattschneider, "Effect of spatial and energy distortions on energy-loss magnetic chiral dichroism measurements: Application to an iron thin film," *Ultramicroscopy,* vol. 110, pp. 1033-1037, 2010.

[36]   J. Rusz, P. M. Oppeneer, H. Lidbaum, S. Rubino, and K. Leifer, "Asymmetry of the two-beam geometry in EMCD experiments," *Journal of Microscopy,* vol. 237, pp. 465-468, 2010.